\documentclass[sigconf,nonacm]{acmart}
\usepackage{balance}
\AtBeginDocument{%
  \providecommand\BibTeX{{%
    \normalfont B\kern-0.5em{\scshape i\kern-0.25em b}\kern-0.8em\TeX}}}

\copyrightyear{2022}
\acmYear{2022}
\setcopyright{acmlicensed}
\acmConference[DDAM '22] {Proceedings of the 1st International Workshop on Deepfake Detection for Audio Multimedia}{October 14, 2022}{Lisboa, Portugal.}
\acmBooktitle{Proceedings of the 1st International Workshop on Deepfake Detection for Audio Multimedia (DDAM '22), October 14, 2022, Lisboa, Portugal}
\acmPrice{15.00}
\acmISBN{978-1-4503-9496-3/22/10}
\acmDOI{10.1145/3552466.3556531}

\begin{document}

\title{Human Perception of Audio Deepfakes}

\author{Nicolas M. M\"uller}
\authornote{These authors contributed equally.}
\affiliation{%
 \institution{Fraunhofer AISEC, TU Munich}
 \streetaddress{Lichtenbergstraße 11}
 \city{Garching near Munich}
 \country{Germany}}
\email{nicolas.mueller@aisec.fraunhofer.de}

\author{Karla Pizzi}
\authornotemark[1]
\affiliation{%
 \institution{Fraunhofer AISEC, TU Munich}
 \streetaddress{Lichtenbergstraße 11}
 \city{Garching near Munich}
 \country{Germany}}
\email{karla.pizzi@aisec.fraunhofer.de}

\author{Jennifer Williams}
\affiliation{%
 \institution{University of Southampton}
 \streetaddress{TODO}
 \city{Southampton}
 \country{United Kingdom}}
\email{j.williams@soton.ac.uk}


\renewcommand{\shortauthors}{Nicolas M. Müller, Karla Pizzi, Jennifer Williams}

\begin{abstract}
The recent emergence of deepfakes has brought manipulated and generated content to the forefront of machine learning research. 
Automatic detection of deepfakes has seen many new machine learning techniques. 
Human detection capabilities, however, are far less explored. 
In this paper, we present results from comparing the abilities of humans and machines for detecting audio deepfakes used to imitate someone's voice. 
For this, we use a web-based application framework formulated as a game. 
Participants were asked to distinguish between real and fake audio samples. 
In our experiment, 472unique users competed against a state-of-the-art AI deepfake detection algorithm for 14912total of rounds of the game. 
We find that humans and deepfake detection algorithms share similar strengths and weaknesses, both struggling to detect certain types of attacks. 
This is in contrast to the superhuman performance of AI in many application areas such as object detection or face recognition. 
Concerning human success factors, we find that IT professionals have no advantage over non-professionals but native speakers have an advantage over non-native speakers. 
Additionally, we find that older participants tend to be more susceptible than younger ones.
These insights may be helpful when designing future cybersecurity training for humans as well as developing better detection algorithms.
\end{abstract}

\maketitle

\section{Introduction}
With the ever-increasing use of voice assistants based on humanized speech, many people are incorporating computer-generated audio into their daily lives.
These voice assistants expose more and more human-like sounds and behaviour \cite{adams2017google,hao2021ai}.
At the same time, generated and manipulated audio data poses a genuine threat to society: 
if people are not able to tell apart which speech is true and which is generated by a machine, then a severe trust issue may follow~\cite{chesney2019deep,westerlund2019emergence}. Our work in this paper helps contextualise the gaps for human deepfake detection performance, and our analysis can be useful for training cybersecurity professionals. 

To the best of our knowledge, we are the first to construct an evaluation paradigm along with a large-scale baseline study to determine whether humans or AI algorithms are currently better at detecting audio deepfakes (i.e., synthetically generated audio). We use the term \textit{deepfake} in this paper to mean speech that was altered or produced based on a deep neural network that can be used (intentionally or unintentionally) to fool a human listener. 
We explore the similarities and differences between human judgements and machine algorithms for detecting audio deepfakes. Our study is also the first to take into consideration whether the humans are native or non-native speakers of the deepfake language and their self-reported IT technical experience. 

The study that we present in this paper is a gamified online experiment where participants compete against an AI algorithm. Previous works have shown that in the naive case, the AI algorithm is exploiting artifacts and shortcuts from the training data~\cite{muller2021speech, geirhos2020shortcut}. Therefore results that compare humans to AI algorithms should be taken with a certain degree of caution. Our contributions are as follows:
\begin{enumerate}
    \item Analyse how well humans distinguish deepfakes from authentic audio and if certain attributes (e.g., native speaker, IT experience, age) affect detection performance. 
    \item Demonstrate that current AI algorithms can out-perform humans when the deepfake audio exhibits certain patterns.
    \item Demonstrate that in realistic scenarios (lacking artificial audio artifacts) AI algorithms perform similarly to humans across many different types of attacks.
    \item Show that humans and machines struggle to detect similar types of deepfakes.
\end{enumerate}
The remainder of this paper describes previous work, explains our experimental setup and design, and provides our analysis of data collected from the online game. 
Finally, we provide some discussion and outlook on this problem.

\section{Related Work}
There are three important aspects of related work: deepfake recognition, synthetic speech evaluation, and spoofing countermeasures. Typically these specializations are treated separately, as spoofing countermeasures commonly refer to countermeasures that protect automatic speaker verification (ASV) systems, rather than deepfake detection in general. Synthetic speech evaluation concerns building tools that can improve deepfake systems so that they become better and sound more natural. 

\subsection{Deepfake recognition}
In recent years, many automatic detection mechanisms for deepfakes have been developed in the field of speech forensics as well as in artificial intelligence \cite{mirsky2021creation}.
However, there is little experimental data on the human understanding of manipulated audio content, especially for manipulated speech.
So far, human deepfake detection has mostly been analyzed in the context of image or video data.
In 2019, Rossler et al.\ compared human and machine detection capabilities \cite{rossler2019faceforensics++}.
Their survey was based on their own new database for deepfake videos and images, divided into three quality levels: original, high-quality deepfake, and low-quality deepfake.
Their findings include that the AI detection algorithm outperforms the human participants, especially when it comes to low-quality image deepfakes. This means at lower resolutions, humans have difficulty distinguishing between real and fake images. The results from \cite{rossler2019faceforensics++} were recently confirmed by \cite{korshunov2020deepfake} and again by \cite{korshunov2021subjective} who further found that both human and machine detection of images can be successfully fooled by deepfakes just by using different generation or synthesis mechanisms.
According to \cite{korshunov2020deepfake}, this is mainly because humans are very consistent in the way they perceive different types of deepfakes. Our experiments in this paper begin to address similar questions for speech audio in order to observe how humans perform detecting several different types of deepfakes. 

The most extensive video deepfake dataset to date was composed and analyzed in \cite{groh2021comparing}. It consisted of three separate online studies with a total of 15,016 participants. Participants were either naive or informed of AI detection predictions. They found that naive human participants and the AI detection algorithms had similar detection accuracy but were fooled by different features. Humans who were informed of the AI algorithm's prediction decisions had improved performance on the task. However, if the AI model predictions were wrong then the human accuracy also decreased.

There is also a large body of related work \cite{chintha2020recurrent,alzantot2019deep,chettri2019ensemble,wang2020densely}, which uses machine learning to identify artifacts in audio waveforms that may be indicative or characteristic of deepfakes.
Such artifacts include noisy glitch, phase mismatch, reverberation, or loss of intelligibility~\cite{wu2012detecting,sahidullah2015comparison}. There are also artifacts that humans do not normally perceive such as various types of extended silence~\cite{muller2021speech}, even though such artifacts are important for the automatic detection of deepfakes from AI algorithms.

\subsection{Evaluation of synthetic speech}
A similar task in speech processing involves using tools to automatically assess the quality of synthetic speech, including text-to-speech (TTS) and voice conversion (VC). For example, large-scale evaluation campaigns take place every two years such as The Blizzard Challenge and the VoiceConversion Challenge \cite{zhou2020blizzard}. Evaluations for this type of speech is often done for naturalness mean-opinion scores (MOS) as rated by human listeners. Naturalness scores follow a Likert scale of 1 to 5 (or less commonly 1 to 10) where 1 is indicative that synthetic speech sounds like a machine and 5 indicates that the speech sounds natural. A similar type of test can be used for assessing pleasantness and intelligibility, though often intelligibility testing asks participants to transcribe text as they heard it and an error rate is calculated by comparing listener transcriptions with ground truth. In all of these types of synthetic speech evaluations, listeners are usually told that they are listening to speech generated by a machine. 

Human ratings, as described above, can be used to develop tools that predict aspects of naturalness automatically. For example, \cite{williams2020comparison} compared different types of speech representations and trained a convolutional neural network to learn a mapping between speech representations and naturalness scores. Similar works have highlighted the inherent difficultly of this task \cite{cooper2022generalization} and a series of more in-depth investigation into automatic MOS-rating tools has gained traction, including a new shared-task challenge \cite{huang2022voicemos}. While rating the naturalness of synthetic speech is related to deepfake detection, these tasks are very different. In the first place, participants do not know which deepfake samples are real or fake and they are asked for a binary decision rather than a gradation. In fact, samples could be rated as a deepfake even though they sound `very natural' and this is what makes human perception of deepfakes so difficult.

\subsection{Spoofing countermeasures}
Deepfakes are often referred to as `generated speech' or `spoofed speech' while authentic speech is often called `bonafide speech'. Recently, audio deepfakes are referred to as \textit{presentation attack}s in the context of applying them to trick an automatic speaker verification system (ASV) \cite{busch2019standards}. The term `countermeasures` refers to devising a protective layer between an ASV system and incoming speech in order to prevent spoofed speech from progressing through an ASV system \cite{wu2015asvspoof}

The international biennial ASVspoof contest\footnote{See \url{https://www.asvspoof.org/}.} challenges researchers to find suitable machine learning algorithms for the detection task.
In \cite{das2020predictions}, the authors used the ASVspoof dataset to see if human-based subjective ratings on \textit{spoofness} can be predicted automatically.
For this, they built an inter-language study with 68 native English and 206 native Japanese listeners classifying samples as spoofed or benign.
They reported that the overall tendencies were the same in both language groups with only minor differences. The authors of the study did not distinguish between those with IT technical experience and those without -- which is something that we address in this paper. A similar human perception test was conducted by \cite{wang2020asvspoof} where participants were asked to consider a scenario where they may work at a call center, and must identify suspicious calls. In that work, the sample size was 1,145 subjects but they did not distinguish between native and non-native speakers or technical expertise. The study that we present in this paper treats the problem as a game where humans play against an AI algorithm.

Recent work from \cite{Wang2021ACS} performed a detailed comparison and analysis of the best-performing countermeasures systems, examining how initial conditions (including random seed) affect countermeasure performance. In fact, \cite{bonastre:hal-03346196} believes that spoofing and countermeasures will perhaps never be a solved problem in computer science. In the meantime, others are working on partial-spoof detection which is a notably more difficult problem for developing ASV spoofing countermeasures \cite{zhang2021initial,zhang2022partialspoof}. 

\section{Experiment}
To compare human and machine detection capabilities, we have designed a game-based challenge. In this section, we describe the structure of the game, the datasets that we used, and the AI algorithm that we used for the machine `player'. We also describe how bonafide and deepfake audio samples were selected to be presented to the human/machine game players. 
\subsection{Experiment description}
\begin{figure}
    \centering
    \includegraphics[width=0.39\textwidth]{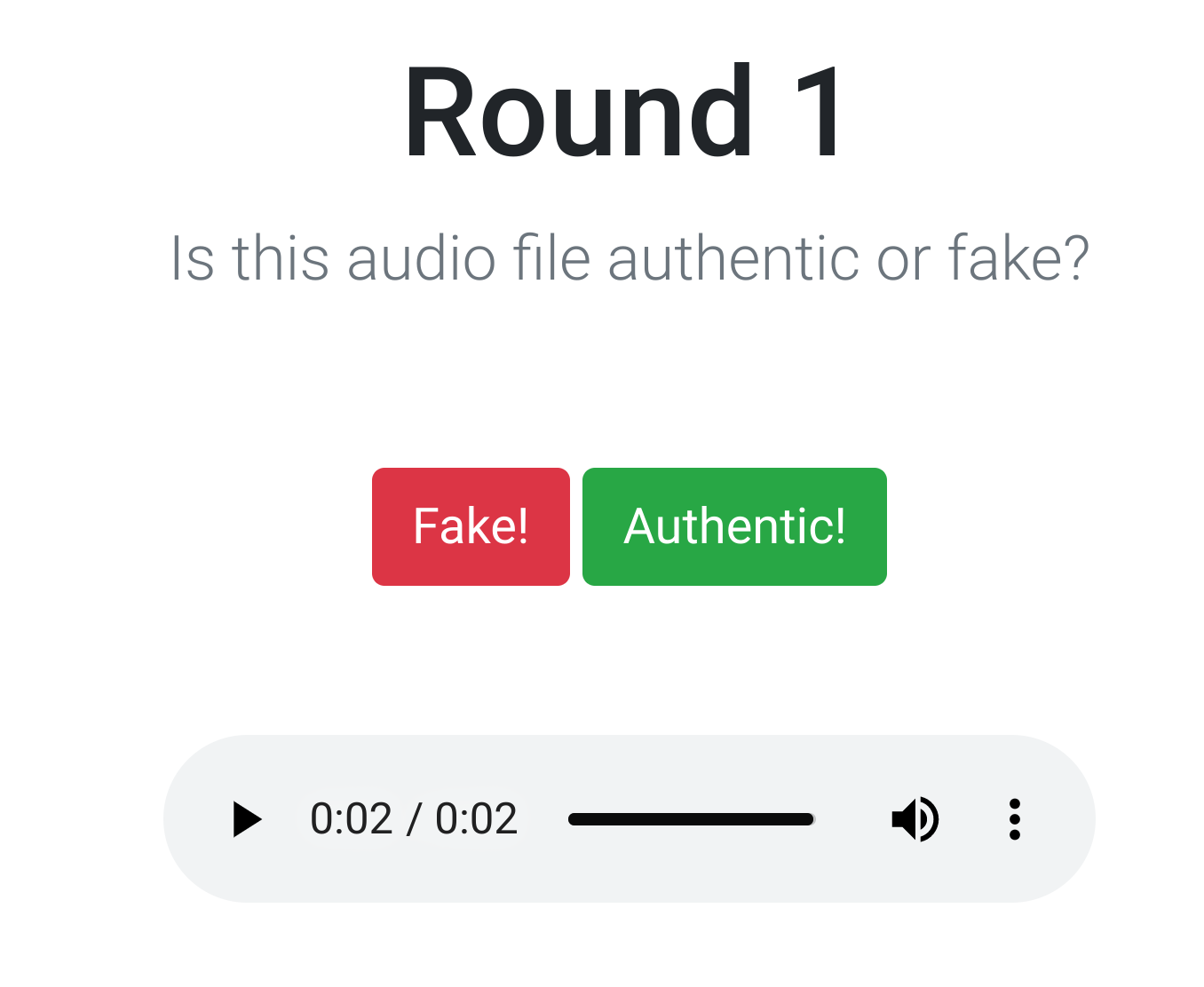}
    \caption{The web interface as presented to the users. 
    The user is required to listen to an audio file (as often as they like) and then classify the audio via the `Fake!' or `Authentic!' button.
    After classifying, the true label is shown to the user along with the AI algorithm prediction for the gamification approach. 
    Available at \url{https://deepfake-total.com/spot_the_deepfake/}.
    }
    \label{fig:interface}
\end{figure}
Our game was distributed online and it forms the basis of our experiments. In our online game, a human player (e.g., a user) competes with an AI algorithm to detect audio deepfakes. The experiment is designed as a gamified classification challenge~\cite{sailer2017gamification}:
the user is provided with one audio file at a time and has to decide whether the audio is authentic (bonafide) or a deepfake, see Figure~\ref{fig:interface}.

Simultaneously, while the user is making a decision a trained machine learning model also classifies the audio file in the background without revealing the prediction.
Once the user has made a decision and classified the audio, we then present the user with the true label along with the AI algorithm's classification prediction.
This way, the user can compare their own detection capabilities to the AI algorithm.
Such a turn is denoted as one `game round'.
The users can play as many rounds as they like.
However, for all subsequent evaluations, we removed users who played less than ten rounds from the dataset. In total, $\protect 472$ individuals have participated in our experiment and each person played at least ten rounds. In summary, $\protect 14912$ game rounds were played.

\begin{figure}
     \centering
         \includegraphics[width=0.49\textwidth]{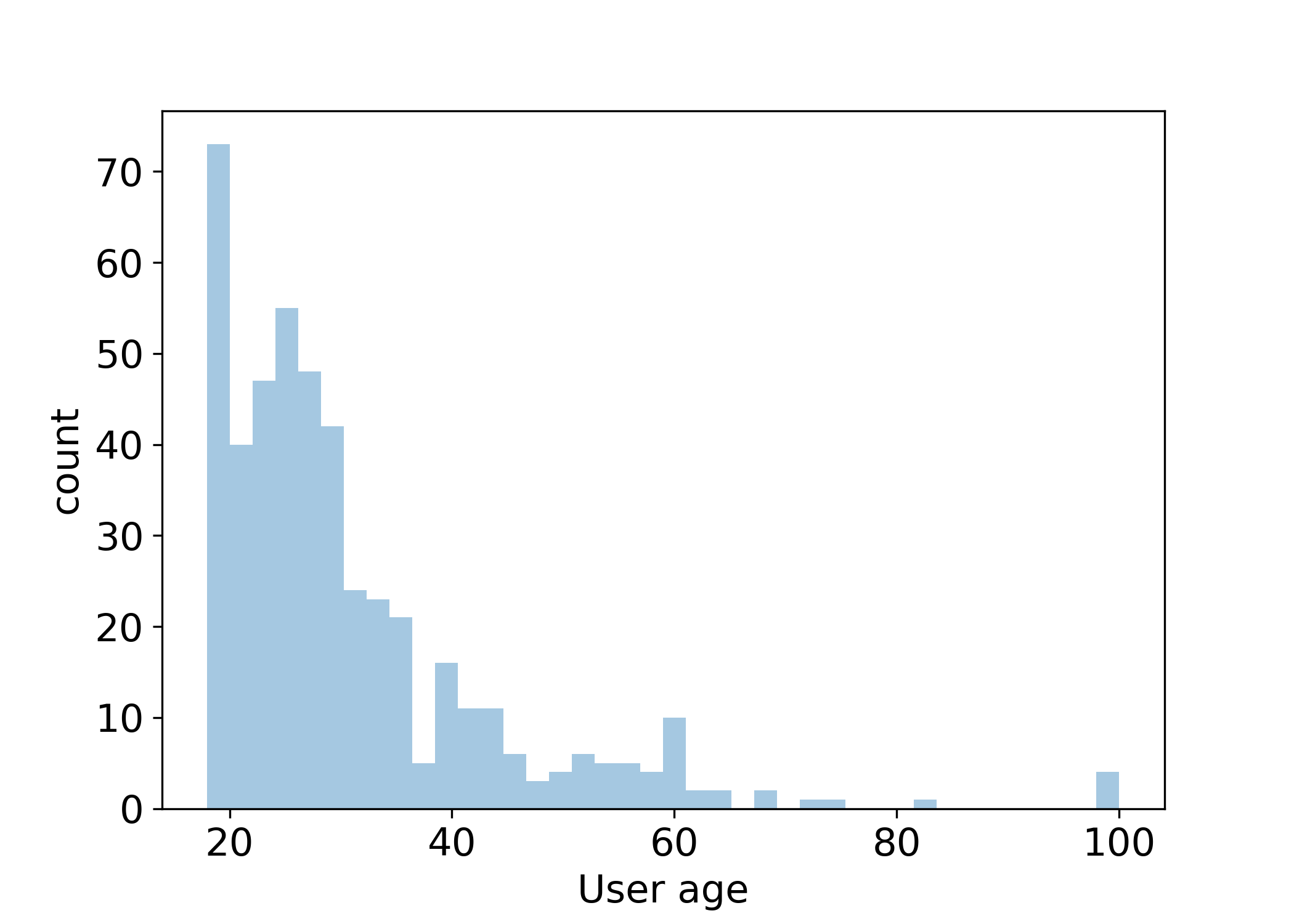}
         \caption{Age distribution of participants who were included in our analysis of the game.}
         \label{fig:ageDistribution}
\end{figure}

To allow for a more detailed analysis, we asked the users to supply some information about themselves. We asked for their IT experience on a Likert scale of 1 to 5 (where 1 is low knowledge and 5 is expert). Users also supplied their age and whether they self-identified as native English speakers.
For our $\protect 472$ participants, the distributions of these characteristics are displayed in 
Table~\ref{tab:skill} and Figure~\ref{fig:ageDistribution}.
There are 105participants for whom English is a mother tongue and 367participants for which it is not.
\begin{table}[]
    \centering
    \begin{tabular}{rr}
\toprule
 Skill Level &  Unique User IDs \\
\midrule
           1 &               32 \\
           2 &               52 \\
           3 &              119 \\
           4 &              158 \\
           5 &              111 \\
\bottomrule
\end{tabular}

    \caption{The number of participants per IT experience level category (1 -- low knowledge; 5 -- expert knowledge).}
    \label{tab:participantsPerIT}
    \label{tab:skill}
\end{table}
\subsection{Audio deepfake dataset}
We base our experiments on the ASVspoof Challenge 2019 dataset \cite{wang2020asvspoof}, a dataset used for bench-marking automatic spoof detection tools and AI algorithms.
It comprises many different types of spoofing attack techniques (referred to as attack IDs). It is also multi-purpose and can be used in tasks such as automatic interactive voice response systems or speaker recognition.
The AI model that we describe in Section~\ref{sec:model} was trained only with the `train' split of the data.
In the online gamified experiment, we present the human players and AI model with audio taken only from the `eval' split of the data. 
Thus, we maintain a proper train/test split.
The `eval' split contains 7355 bonafide audio samples. For each of the 13 different spoofing attack techniques in the evaluation dataset (`A07' - `A19'), there are 4919 spoofed audio samples.
The authors employ a set of spoofing systems, where each model is categorized either as a Text-to-Speech model (TTS), voice-conversion model (VC) or waveform synthesis model (WC), c.f.~\cite{wang2020asvspoof}.
\subsection{Model selection and training}\label{sec:model}
We trained and evaluated two different machine learning models. First, we implemented a baseline model which is similar to the models used in related work~\cite{lavrentyevaaudio2017,lavrentyevastc2019, wang2021Comparative}. This baseline model is a three-layer bi-directional LSTM (BiLSTM) consisting of 256 hidden neurons and it was trained with 10\% percent dropout. For this model, the audio is converted to spectrograms. 

For the second model, we implemented RawNet2~\cite{rawent2} which a state-of-the-art model known to perform well on audio deepfake detection in the ASVSpoof challenges.
The RawNet2 architecture uses raw waveforms as input and processes the waveforms using the SincNet architecture \cite{ravanelli2018speaker}, residual blocks, and gated recurrent units (GRUs).
Following~\cite{muller2021speech}, we removed the leading and trailing silence from the input audio. It was established by \cite{muller2021speech} that simply the length of the audio silence behaves as a learning shortcut~\cite{geirhos2020shortcut} in the ASVspoof 2019 challenge training data by correlating strongly with ground truth attack class labels.

Thus, our RawNet2 architecture represents a state-of-the-art model, tested under realistic conditions.
Both of these models yield a prediction score $\lambda \in \mathbb{R}$, which requires a threshold to perform binary classification.
We continuously compute this threshold over the previously seen instances in the game.
Doing so allows us to represent the fact that humans learn individually while recognizing that the trained AI algorithm has the potential to observe all game rounds seen so far regardless of which human user is playing.
\subsection{Selection of queried instances}
We select the individual audio samples to be played to the user in the game with equal probability such that the bonafide and spoofed audio files were each sampled with a probability of $p=0.5$.
Inspired by active learning~\cite{felder2009active}, we over-sample instances that are more difficult human players. For example, during the game if we observe that attack $A10$ is more difficult for players then that attack will be over-sampled compared to attacks that are easier.
Specifically, when sampling spoofed audio, we first sample (weighted with weights $w_i$) the attack ID $i \in \{$A07$, \, $A08$, \, ..., $A19$\}$, and then randomly select audio from within an attack IDs based on a uniform distribution.
The attack IDs are sampled using the weights in Equation 1:
\begin{equation}
    w_i = 1 - \frac{\text{acc}_{i}}{1+\epsilon}
\end{equation}
\noindent where $\epsilon = 0.03 > 0$ to assert $w_i > 0$. The term $\text{acc}_{i} \in [0, 1]$ denotes the overall human accuracy for samples created by attack ID $i$.
Hence, attacks that are more difficult for other humans to spot are more likely to be used in the current game round.
The rationale behind this design choice is that there are several attacks in the dataset that are trivial to spot. 
Playing these repeatedly to the user would yield very little information (see attack $A13$ in Figure~\ref{fig:accuracyPerAttack}, where the human players have an accuracy of $> 0.99$).
\section{Results}\label{sec:results}
\begin{figure}
     \includegraphics[width=0.49\textwidth]{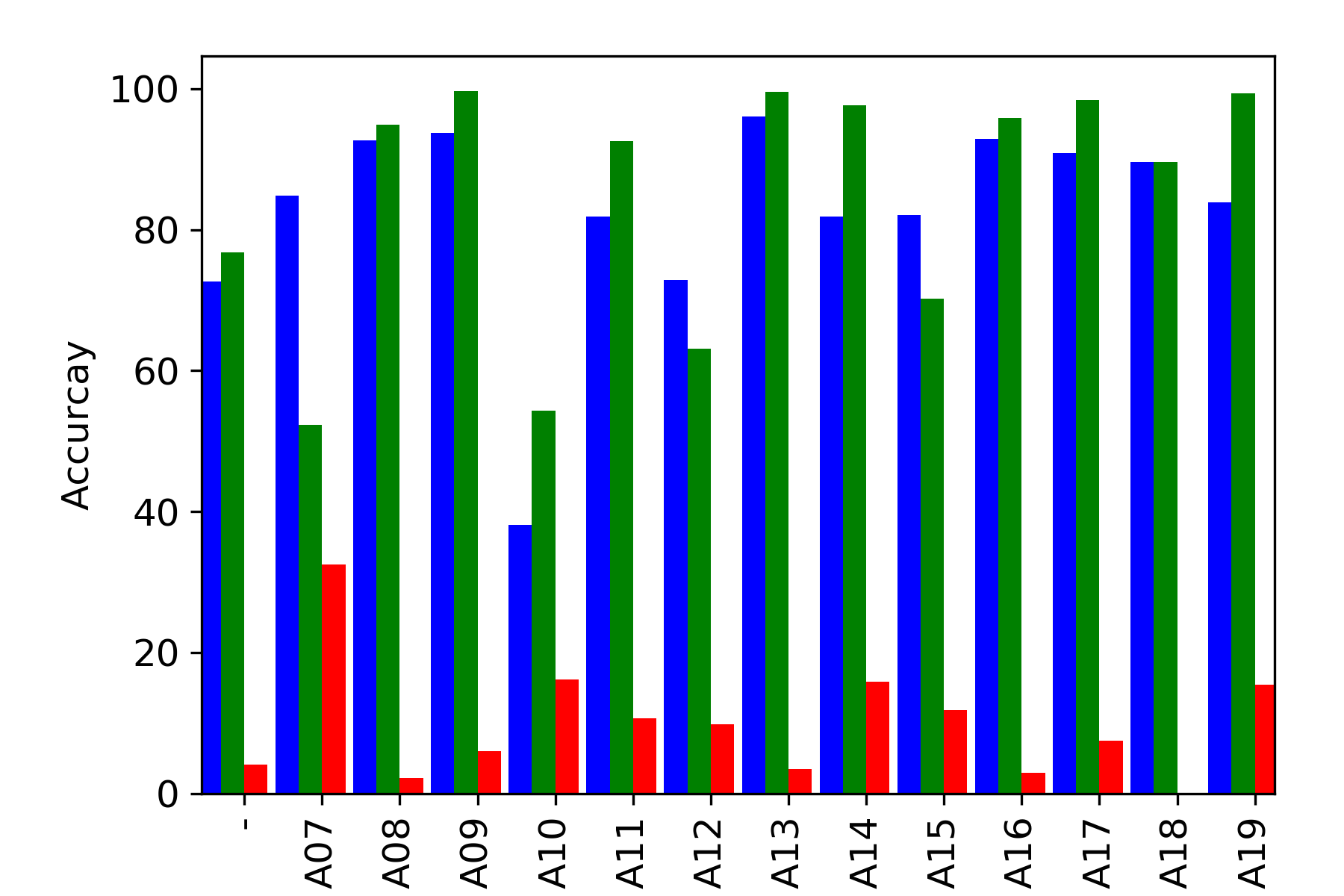}
     \caption{\textbf{Average accuracy per attack}.
     This graph shows the mean attack accuracy per attack ID (with `-' denoting no attack, i.e., bonafide samples).
     The blue bars indicate the accuracy of the human players, the green bars indicate the accuracy of the AI (RawNet2).
     The absolute difference is shown in red.
     The differences between human players and the AI algorithm is small at \protect\input{res/auto_created/mean_delta_human_ai}, on average.}
    \label{fig:accuracyPerAttack}
\end{figure}

\begin{figure}
    \centering
    \includegraphics[width=0.45\textwidth]{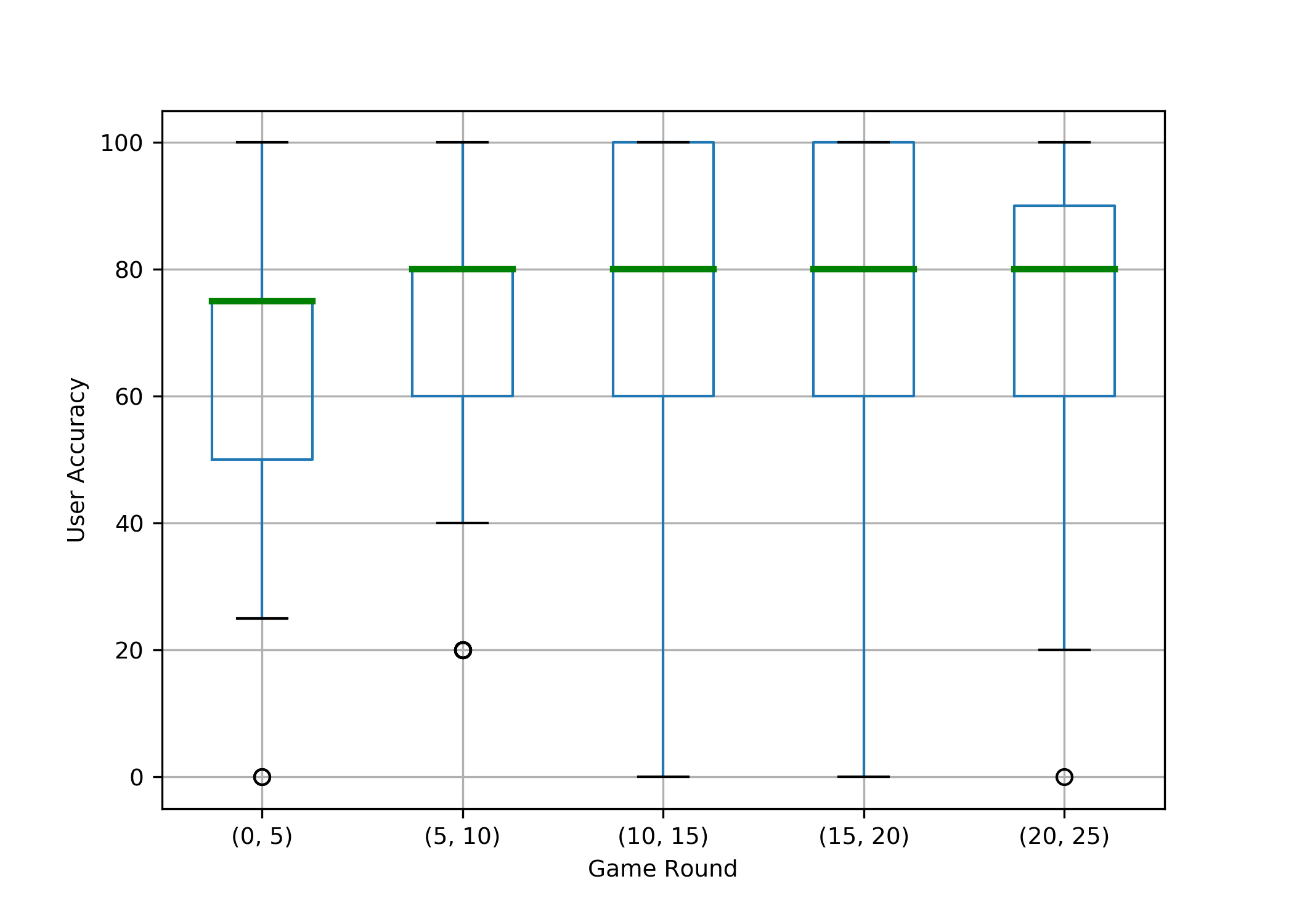}
    \caption{User accuracy \textbf{grouped by number of game rounds played}. This figure shows human detection accuracy ($y$ axis) for the first ten game rounds, second ten game rounds, etc. (intervals shown on $x$ axis). Note that after a few warm-up rounds (i.e., the first $20$ rounds), detection accuracy plateaus.} 
    \label{fig:accuracyPerGameRound}
\end{figure}

In this section, we present the results of our experiment considering the performance of humans vs. machine, the correlation between accuracy and user characteristics, and the eventual plateau of human performance. Note that we exclude users who played less than 10 rounds, which leaves $\protect 472$ participants who played a total of $\protect 14912$ rounds. The results for each main finding will be presented in this section using  Figures~\ref{fig:accuracyPerAttack} to \ref{fig:accuracyPerGameRound}. Here we state a summary of the key findings as follows: 
\begin{enumerate}
    \itemsep0em
    \item The naive AI-based approach overfits to artifacts in the dataset and, hence, outperforms the human players.
    \item In a realistic scenario, our state-of-the-art AI algorithm performs similarly to the human test subjects, where both struggle with the same types of attacks.
    \item TTS-based spoofing systems fool humans much more consistently than voice-conversion or waveform concatenation systems.
    \item 
    We find that a) native speakers have an advantage in deepfake detection, 
    b) the self-reported IT experience and skill level for human players does not impact the detection rate, and
    c) human detection performance deteriorates with age.
    \item The ability of humans to detect deepfakes improves with training during the first few game rounds. 
    However, it quickly plateaus.
\end{enumerate}

\subsection{Naive AI vs. human participants}
AI superiority in detecting certain types of deepfakes has already been shown by Groh et al.~\cite{groh2021comparing} 
and others \cite{rossler2019faceforensics++,korshunov2020deepfake,korshunov2021subjective}.
We can verify this for our naively constructed scenario, which uses a learning shortcut~\cite{muller2021speech,geirhos2020shortcut} in the ASVSpoof 2019 dataset. A learning shortcut allows a model to solve a classification task using a set of trivial features, such as the presence of silence (equals authentic audio) or absence of silence (equals spoof) in an audio file. Such shortcuts should -- if possible -- be removed from the data because they prevent true model  generalization. 
In our experiment scenario, the AI algorithm achieves 95\% accuracy on ASVspoof 2019, while humans only obtain 80\%.

\subsection{Realistic AI vs. human participants}
However, under more realistic conditions (i.e., after the shortcut removal), AI and humans perform largely similarly.
In particular, they share the same strengths and weaknesses.
Consider Figure~\ref{fig:accuracyPerAttack}, a comparison of the machine and human detecting capabilities per attack: the attacks $A10$ and $A12$ (both advanced TTS systems, namely Tacotron 2~\cite{shen2018natural} and Wavenet~\cite{oord2016wavenet}) are difficult for both the AI algorithm and humans (about 50-60\% accuracy).
$A09$, $A13$, and $A16$ (TTS, VC and WC systems, respectively) are easy to detect for both AI and humans (more than 95\% accuracy).
Interestingly, attack $A07$ (a TTS system) is challenging for the AI, but not so much for humans.
This may be due to $A07$ being the only attack to employ a Generative Adversarial Networks (GAN) as a waveform generator.
In such models, an adversarially trained discriminator tries to remove all machine-perceptible traces of the audio's spoofness; thereby possibly limiting the performance of subsequently applied audio-spoof detection models.
We invite the reader to listen to samples at: \url{https://deepfake-total.com/spot_the_deepfake}.

\subsection{Comparison of audio spoof architectures}
\begin{figure}
    \centering
    \includegraphics[width=0.45\textwidth]{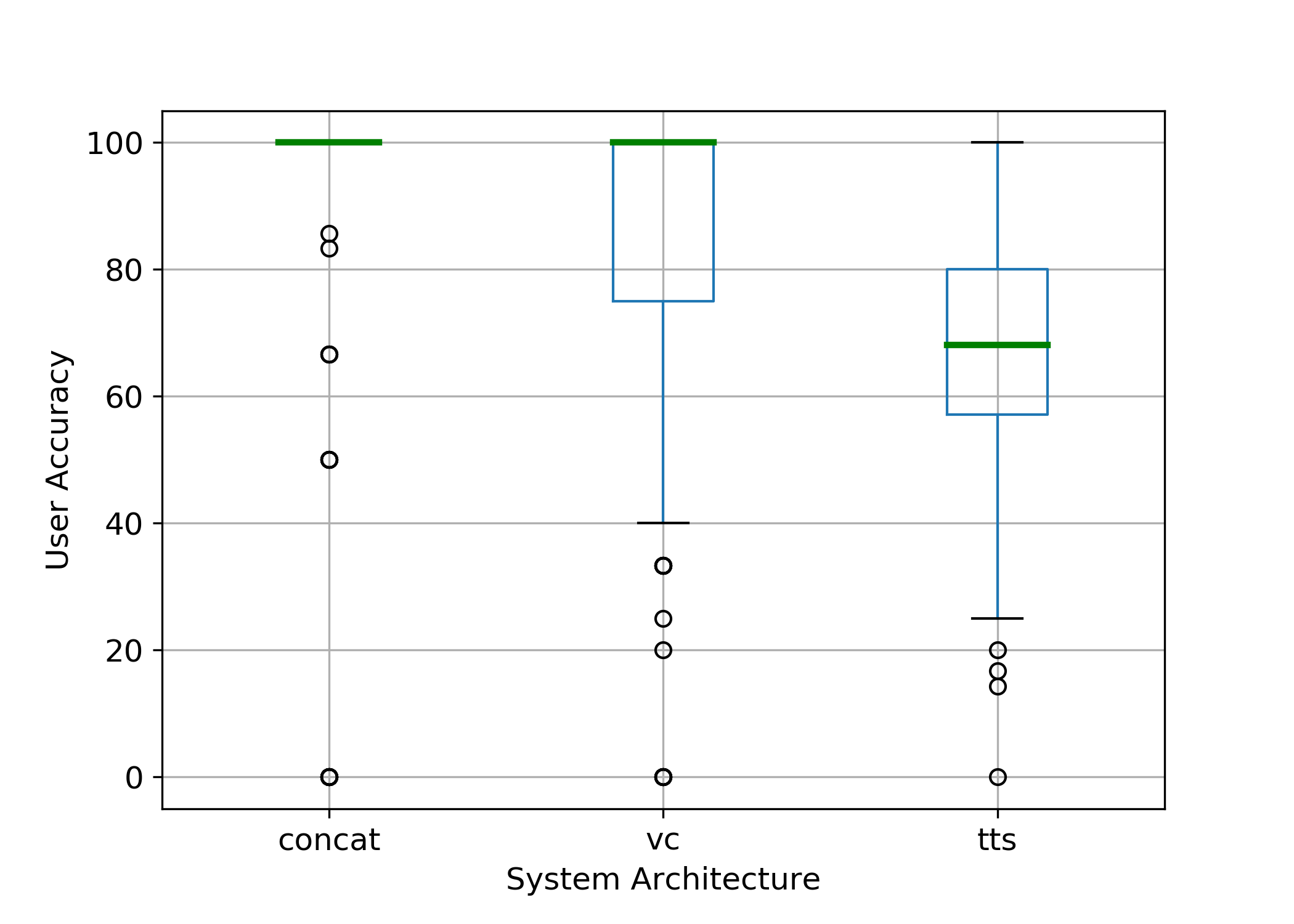}
    \caption{Mean user accuracy vs. system architecture of the audio spoof system. TTS-based systems (tts) clearly outperform voice conversion (vc) and waveform concatenation (concat).}
    \label{fig:sys_archi}
\end{figure}

Consider Figure~\ref{fig:sys_archi}, which shows the average human detection performance vs. the type of ASVspoof attack architecture, derived from~\cite{wang2020asvspoof} for each attack in the `eval' data, i.e., attacks $A07$ -- $A19$.
Text-to-speech systems (TTS) fool human participants much better than other systems (voice conversion and waveform concatenation).

\subsection{Self-reported user characteristics}
In our experiment, we queried the user's age, self-reported IT-experience, and whether they have English as a mother tongue.
Here, we analyze if any of these factors correlates with detection performance.

\subsubsection{IT-experience}
\begin{figure}
    \centering
    \includegraphics[width=0.45\textwidth]{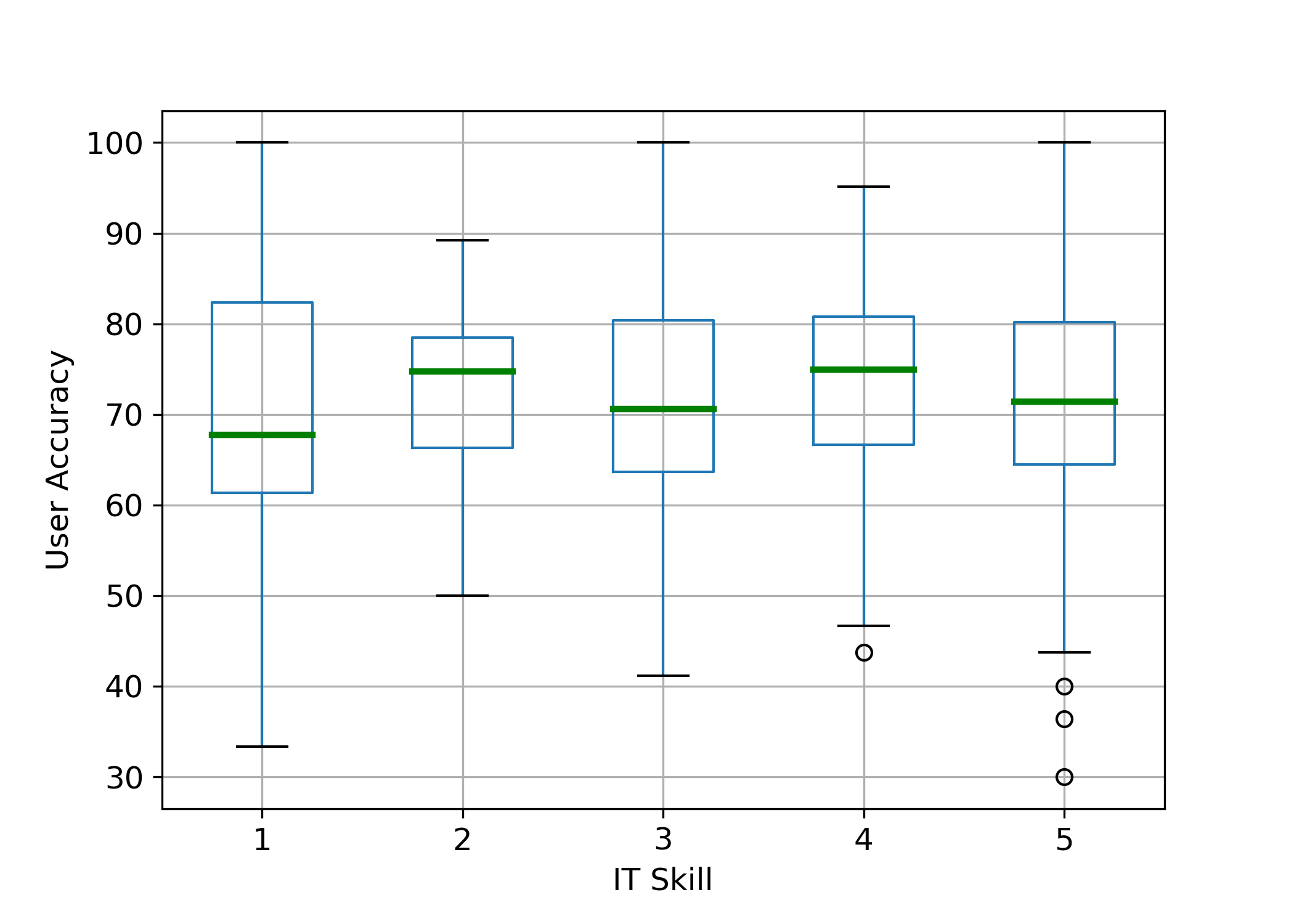}
    \caption{Human detection accuracy \textbf{grouped by the level of IT expertise} (1 -- little knowledge; 5 -- expert knowledge).
    There is no significant correlation between the level of expertise and the ability to detect audio deepfakes.}
    \label{fig:accuracyPerExperienceLevel}
\end{figure}

We observe that there is no correlation between self-reported IT experience levels and the ability to detect audio deepfakes, c.f.~Figure~\ref{fig:accuracyPerExperienceLevel}.

\subsubsection{Native Tongue}
\begin{figure}[t]
     \centering
         \includegraphics[width=0.45\textwidth]{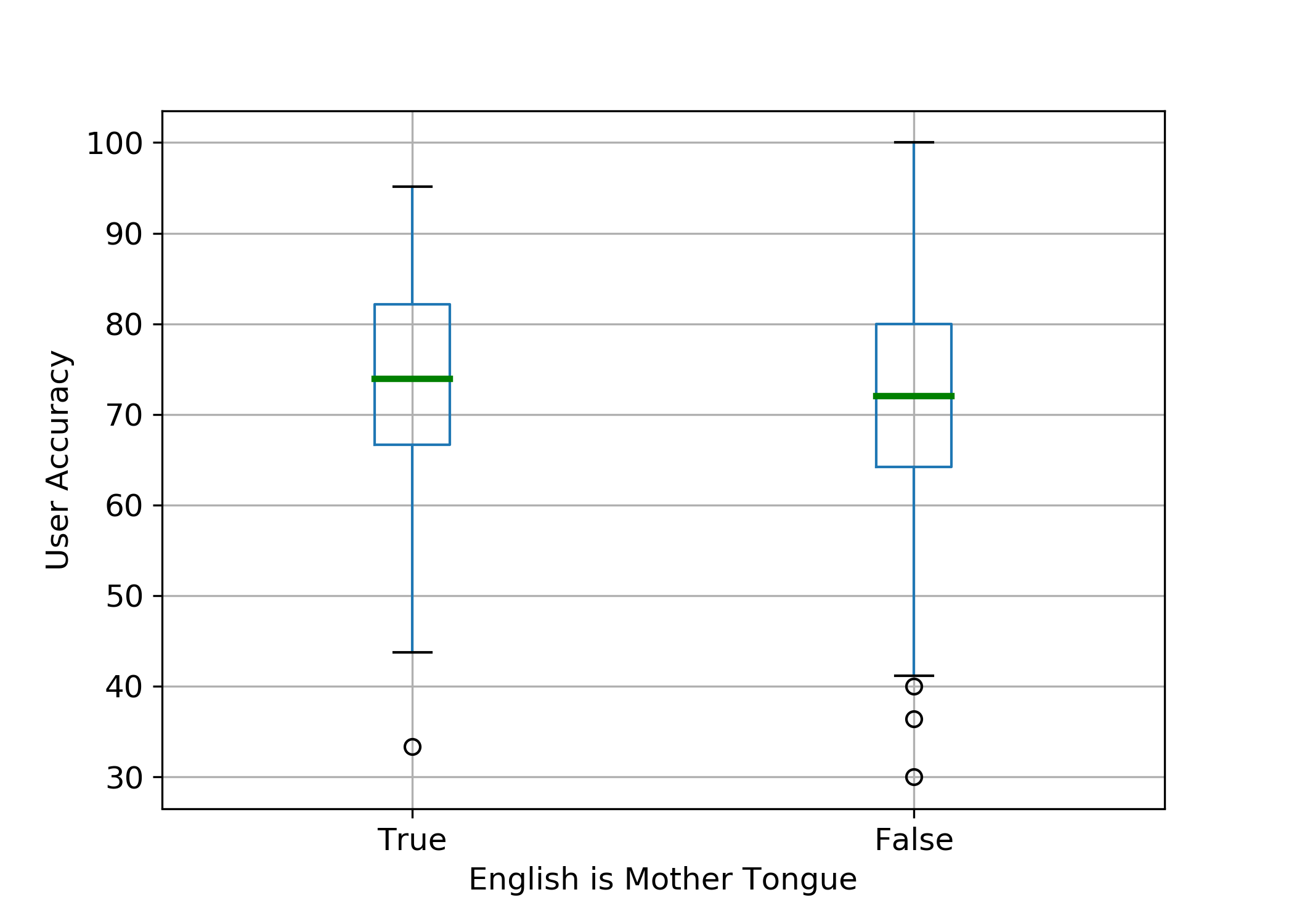}
         \caption{Deepfake detection accuracy of users grouped by whether or not they are \textbf{native English speakers}.
         Native speakers perform slightly better in deepfake detection.
        }
         \label{fig:MotherTongue}
\end{figure}

Some improvement in the recognition rate can be seen when comparing native speakers to non-native speakers. 
People with English as a mother tongue can recognize audio deepfakes slightly points better on average, as shown in Figure~\ref{fig:MotherTongue}.

\subsubsection{Age}
\begin{figure}
    \centering
    \includegraphics[width=0.45\textwidth]{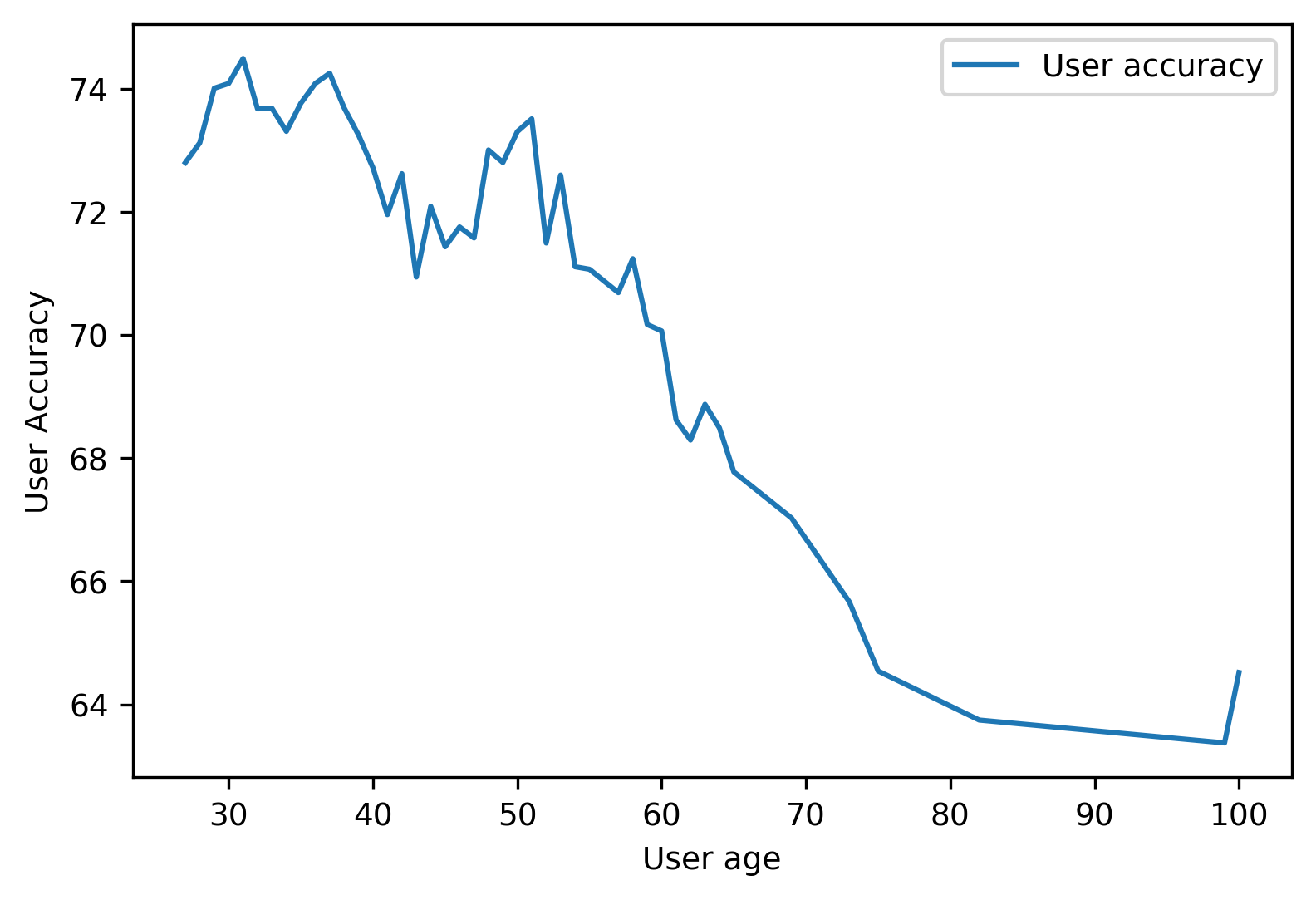}
    \caption{Mean user accuracy vs. age; aggregated over a rolling window of ten years. Detection performance declines with increasing age.}
    \label{fig:detection_vs_age_rolling}
\end{figure}

Consider Figure~\ref{fig:detection_vs_age_rolling}, which shows the average human detection performance, averaged per age and smoothed via a rolling window over $n=10$ years.
We see that detection performance steadily declines, indicating that older subjects may be more vulnerable to audio deepfakes than younger ones.
We presume this is because younger participants tend to be 'digital natives', and the increased exposure to digital content improves their judgement.
Alternatively, age-related hearing loss \cite{decibel2021high} could be a deciding factor, causing older people to be less sensitive to manipulated high frequencies, which are often characteristic of synthetic speech.

\subsection{Performance vs. human exposure}
How does exposure to audio deepfake and learning feedback from the system impact a user's performance?
In order to find out, we proceed as follow:
First, we filter our results for users who played at least 20 game rounds.
We then compare their performance within different game rounds.
We observe that detection accuracy improves after the first 10 rounds from $67\%$ to $80\%$.
However, more training or exposure to deepfakes, along with feedback stating the correct label `spoof' or `benign' after each game round, does not appear to help users improve: the accuracy quickly stagnates at $80\%$.

\section{Discussion and Outlook}
We find that state-of-the-art audio deepfake detection models lack the super-human performance observed in other domains such as image and video.
Deepfake detection algorithms perform similarly to human test subjects, sharing the same strengths and weaknesses, at least under realistic settings. 
It is important to consider that native English speakers perform better at detecting audio deepfakes that consisted of English speech. 
We expect that the effect of native language holds across multiple languages, though this is an interesting avenue for future work. 
We use English audio samples because audio deepfake data is readily available in English. 
Additionally, this makes our online game most accessible.
However, we encourage the creation of multilingual deepfake datasets and suggest that this is incorporated into future ASVspoof challenges.  

The creation of audio deepfakes itself is a young discipline: the first text-to-speech (TTS) synthesis models using neural vocoders were published in 2017~\cite{wang2017tacotron}. So it is not surprising that audio deepfake detection is still in an early stage.
We hope that future research might further improve detection performance as well as develop more training programs for humans.
This is highly needed because deepfakes (both voice conversion and TTS synthesis) will continue to improve over time, while we expect the human detection rate to find a plateau. With all of these considerations, improved countermeasures will be increasingly necessary. 

\section{Acknowledgments}
This research was supported by the Bavarian Ministry of Economic Affairs, Regional Development, and Energy as well as the German Federal Ministry for the Environment, Nature Conservation, Nuclear Safety and Consumer Protection.

\bibliographystyle{unsrtnat} 
\balance
\bibliography{mybib}


\end{document}